# Disagreeing about Crocs and socks: Creating profoundly ambiguous color displays

Pascal Wallisch & Michael Karlovich

*Department of Psychology and Center for Neural Science, New York University, 4 Washington Place, New York, NY 10003, USA.*

Corresponding author: Pascal Wallisch - pascal.wallisch@nyu.edu - @pascallisch

Abstract: There is an increasing interest in the systematic disagreement about profoundly ambiguous stimuli in the color domain. However, this research has been hobbled by the fact that we could not create such stimuli at will. Here, we describe a design principle that allows the creation of such stimuli and apply this principle to create one such stimulus set - "the crocs and socks". Using this set, we probed the color perception of a large sample of observers, showing that these stimuli are indeed categorically ambiguous and that we can predict the percept from fabric priors resulting from experience. We also relate the perception of these crocs to other color-ambiguous stimuli - "the dress" and "the sneaker" and conclude that differential priors likely underlie polarized disagreement in cognition more generally.

Keywords: Disagreement, Dress, Color, Perception, Ambiguity, Uncertainty, Priors

Acknowledgments: We would like to thank Cecilia Bleasdale for taking the original dress picture and anyone who either helped us to get the data, in particular Cecilia Hua and anyone else who donated their data, shared the link or both.

If you want to contribute your data to a followup project, you can do so here:
https://www.surveymonkey.com/r/crocPerception

Author contributions: PW & MK conceived and designed the study. MK & PW created the stimuli. PW designed the survey. PW & MK recorded the data. MK analyzed the stimuli. PW analyzed the data. PW & MK wrote the manuscript.

Competing financial interests and commercial relationships: The authors declare no competing financial interests, nor any commercial relationships. In particular, we want to proactively affirm that there was no coordination or communication of any kind with Crocs Inc. or any affiliate, regarding any aspects of this study. We chose our stimulus materials solely on the basis of color considerations. We also would like to emphasize that this study was entirely self-funded.

**Introduction**

Strong perceptual illusions - systematic mismatches between the conditions in the external world and subjective experience - have been known for a long time. Notoriously, prolonged exposure to a moving stimulus leads to the vivid subjective experience of motion in the opposite direction once the stimulus objectively stopped moving, a phenomenon known as the motion aftereffect (Purkinje, 1825). Similar phenomena exist in the color domain - for instance, the perceived hue of a colored patch can be shifted dramatically from what it would appear in isolation, if presented with a suitably patterned background (Shevell & Monnier, 2003) or under carefully modulated lighting conditions (Lotto & Purves, 1999), suggesting that observers take both background and illumination into account when judging the appearance of a colored patch. Importantly, all of these illusions are unidirectional, as they seem to affect all observers similarly, predictably shifting the percept in the same direction.

This state of affairs changed in February 2015, when a categorically ambiguous stimulus in the color domain surfaced. Most observers perceived this stimulus - the photo of a wedding dress - as either black and blue (the appearance of the physical dress under sunlight) or as white and gold, which could not be attributed to problems with fundamental color vision or a misapplication of color labels (Lafer-Sousa et al., 2015). The reason this stimulus - "the dress" - is considered categorically (or profoundly) ambiguous is owed to the fact that different observers perceive different kinds - not degrees - of hue, and these hues are on opposite sides of the color wheel. Notably, this is not a bi-stable stimulus - only a few observers (on the order of 1%) - experience a percept that switches back and forth, akin to genuinely bistable stimuli like the Necker cube (Rubin, 1921), binocular rivalry (Levelt, 1968) or moving plaids (Hupe & Rubin, 2003), all with characteristic and known perceptual dynamics. Instead, most observers seem to be "locked in" to a particular interpretation of the image, consistent with the notion that accumulated experience - in the form of an illumination prior - drives the perception of "the dress" stimulus (Witzel et al., 2017). Indeed, the specific kind of life experience that could shape the formation of differential lighting priors, such as differing degrees of exposure to sunlight predicts both the perception of "the dress" stimulus as well as the readiness of observers to assume particular kinds of illuminations of the "the dress" (Wallisch, 2017a). However, while being able to explain the perception of a specific stimulus display in a coherent and meaningful way is compelling, it is also possible that we just managed to shoehorn an "explanation" of a very specific stimulus, without being able to generalize. It would be reassuring if the principles invoked in the explanation of such stimuli allowed us to create novel ambiguous displays at will. Should this succeed, we could have more confidence that it is really these principles governing the perception of such displays. Alas - to date - all other categorically color ambiguous display that have surfaced since "the dress", including the Adidas Jacket (Wallisch, 2016), the flip-flops (Wallisch, 2017b) or the sneaker (Wallisch, 2017c) were not generated purposefully with these principles in mind, but rather arose spontaneously on the internet, allowing for residual doubt whether we truly understand this phenomenon to persist. Here, we propose and use principles that allow for the creation of categorically ambiguous color displays and explore how differential priors determine the subjective perceptual experience evoked by these displays in human observers.

**Method**

*Stimulus design*

We designed our stimuli according to the following principle, guided by the notion that patterned backgrounds and lighting conditions powerfully influence the perception of a color display. Specifically, we first maximize uncertainty by using a monochromatic object that could - in principle be almost any color. Here, we used two versions of "Classic crocs™", namely "Ballerina pink" and "New mint", out of the 28 different colors on offer. We then put these crocs on an unpatterned, black background to make the display devoid of visual cues which could allow observers to calibrate their percept (Land & McCann, 1971). Next, we carefully titrated complementary light to equate the light reflected off the croc to shades of grey.

A final step relies on another piece of garment - in this case the white socks - to mirror the properties of the illumination source. Thus, we introduce the concepts of a colored target object (TCO) - this is the percept that we will ask our objects to judge, corresponding in this case to the color of the crocs. We also introduce the concept of an anchoring or cueing object, which reflects the color of the illumination, hence illumination mirroring object (IMO), in this case the socks, see figure 1.

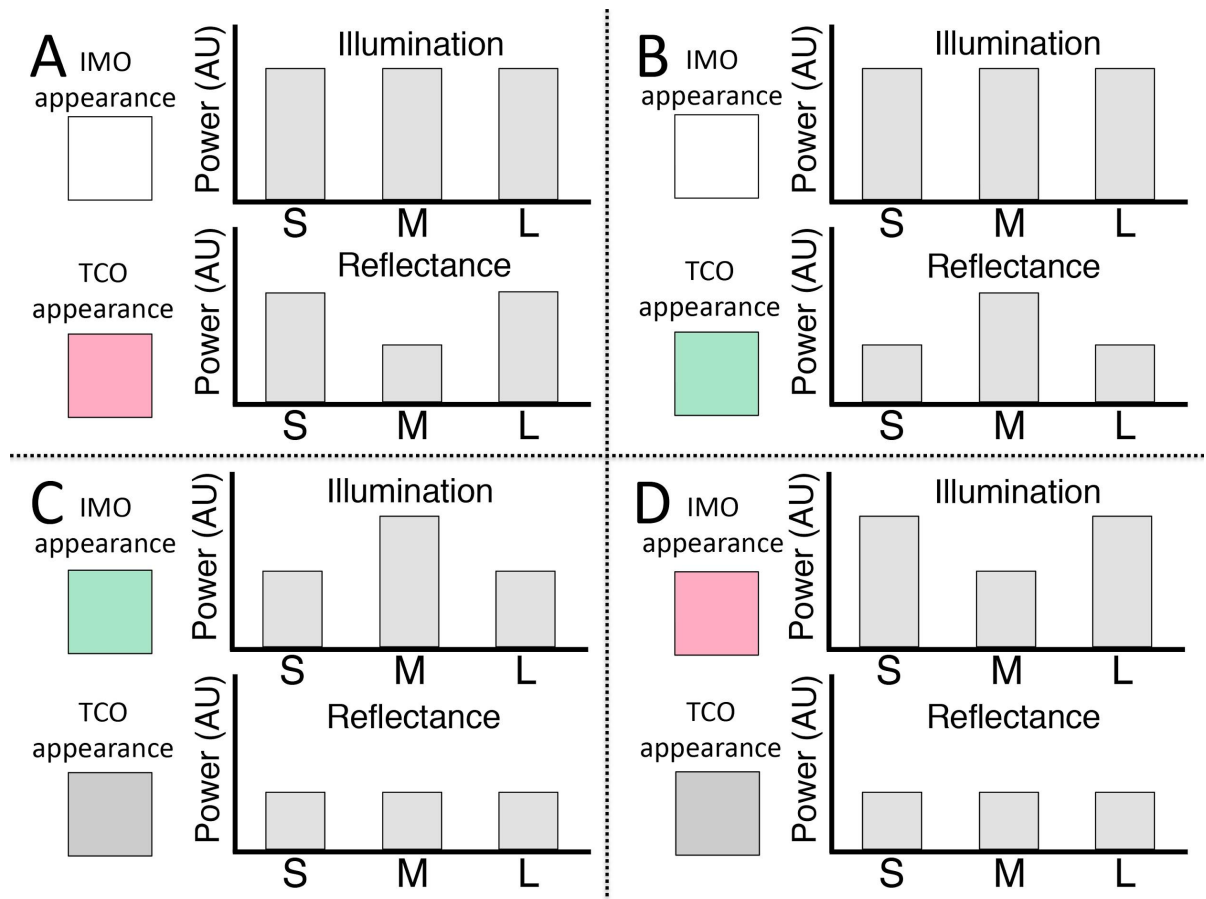

Figure 1: A schematic of the stimulus design principles. Top row: colored target objects (crocs) under typical lighting conditions. We define sunlight or bright indoor light (e.g. incandescent or LED) as "typical" lighting conditions, as these are conditions observers can be expected to experience in everyday living. Bottom row: The same crocs under altered lighting conditions,

carefully titrated to equalize the reflectance. In this view, the crocs can be conceived of as a filter. Left panel: Pink crocs as the target object. Right panel: Mint crocs as the target object. Note that this schematic is highly stylized, with arbitrary units of power (AU) on the y-axis. In reality, the power of the illumination is continuous, not discrete. In addition, the power of "white" light is actually not equal across wavelengths; wavelengths in the middle of the visible range are perceived as much brighter by human observers, due to human spectral sensitivity (Wald, 1964). A: Pink crocs, illuminated by white light appear pink because the croc filters wavelengths in the middle of the range. B: Mint crocs, illuminated by white light appear mint green because the croc filters wavelengths in the short and long range. C: Pink crocs (the CTO), illuminated by light with more power in the middle range, as reflected by the appearance of the IMO, appear as some shade of grey. D: Mint crocs (the CTO), illuminated by light with more power in the short and long range, as reflected by the appearance of the IMO appear as some shade of grey. Note that in all cases, the reflectance of the white sock corresponds to the illumination, mirroring the properties of the light.

This part of the display - the IMO - is also subtly color ambiguous. Whereas this kind of plain tube sock is likely to be white - as they usually are, when experienced in everyday life - the person could possibly be wearing colored socks, in which case the socks provide no information about the lighting. In other words, we use an object that could be any color, but usually isn't - like plain tube socks to serve this mirroring function to create a conflict between color appearance and typically experienced color. We predict that some people - those who have a strong prior that socks like these are white - will use this cue to disambiguate the display and discount the illuminant, but others - those who don't perceive the socks as white - to take the appearance of the socks at face value. This is in line with other research suggesting that "priors" - expectations and assumptions about the properties of stimuli that are typically derived from experience - play a central role in cognition (Körding & Wolpert, 2004; Stocker & Simoncelli, 2006; Griffiths et al., 2008). To summarize, we predict that these principles will yield a categorically color-ambiguous display. We conceive of this design principle as "Substantial Uncertainty combined with Ramified or Forked Priors and Assumptions yields Disagreement" (SURFPAD). We used the following seven images of crocs and socks as stimuli in this research, see figure 2.

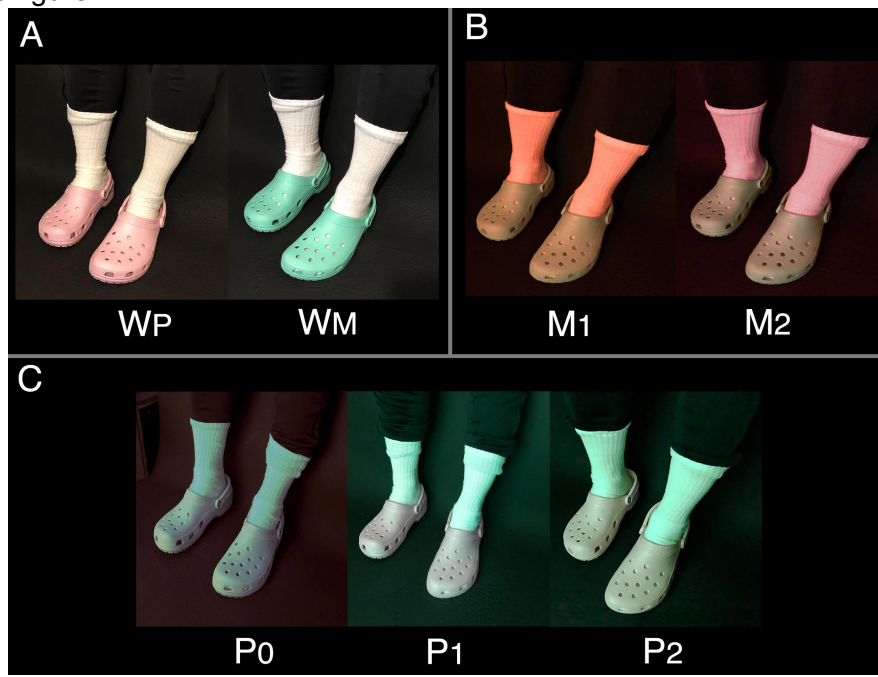

Figure 2: The crocs and socks images we created. A: crocs and Socks under typical (a combination of daylight coming through a window and a LED (Feit electric, 60 Watt equivalent, 5000K, 800 Lumen) lighting conditions. $W_P$: "Ballerina pink" crocs. $W_M$: "New mint" crocs. B: The "New mint" crocs under colored, specifically pink/orange light, yielding two specific stimuli with distinct properties, as reflected by the socks, $M_1$ and $M_2$. C: The "Ballerina pink" crocs under colored, specifically green light, yielding three specific stimuli with distinct properties, as reflected by the socks, namely $P_0$, $P_1$ and $P_2$. The lighting conditions in B and C were created via a dimmable SMD 5050 LED light strip under blackout conditions and on black LifeSpan® treadmill matting. Note that the granularity of the dimming settings of the light strip in combination with the reflectance properties of the crocs (e.g. the "New mint" version was reflecting more strongly in the short wavelengths than anticipated) did not always allow us to completely "whiten" the light reflected off the crocs. In those cases, we applied minimal digital filtering with Adobe® Photoshop® CC 2019 software to close the distance. However, we strongly believe that the use of a more sophisticated lighting system that allows for a sufficiently fine titration of the different light channels would obviate the need for any digital post-production. We first created $P_0$ as a proof of concept. We then created two sets of pairs of stimuli using both pink and mint crocs - and the same socks - under different lighting conditions, P1/P2 and M1/M2, respectively.

We combined these croc displays we created with two others - "the dress" and "the sneaker", both recognized as categorically ambiguous color displays, to use as stimuli in our study. Figure 3 shows the entire set, in combination with gamuts of the individual pixels in hsv space.

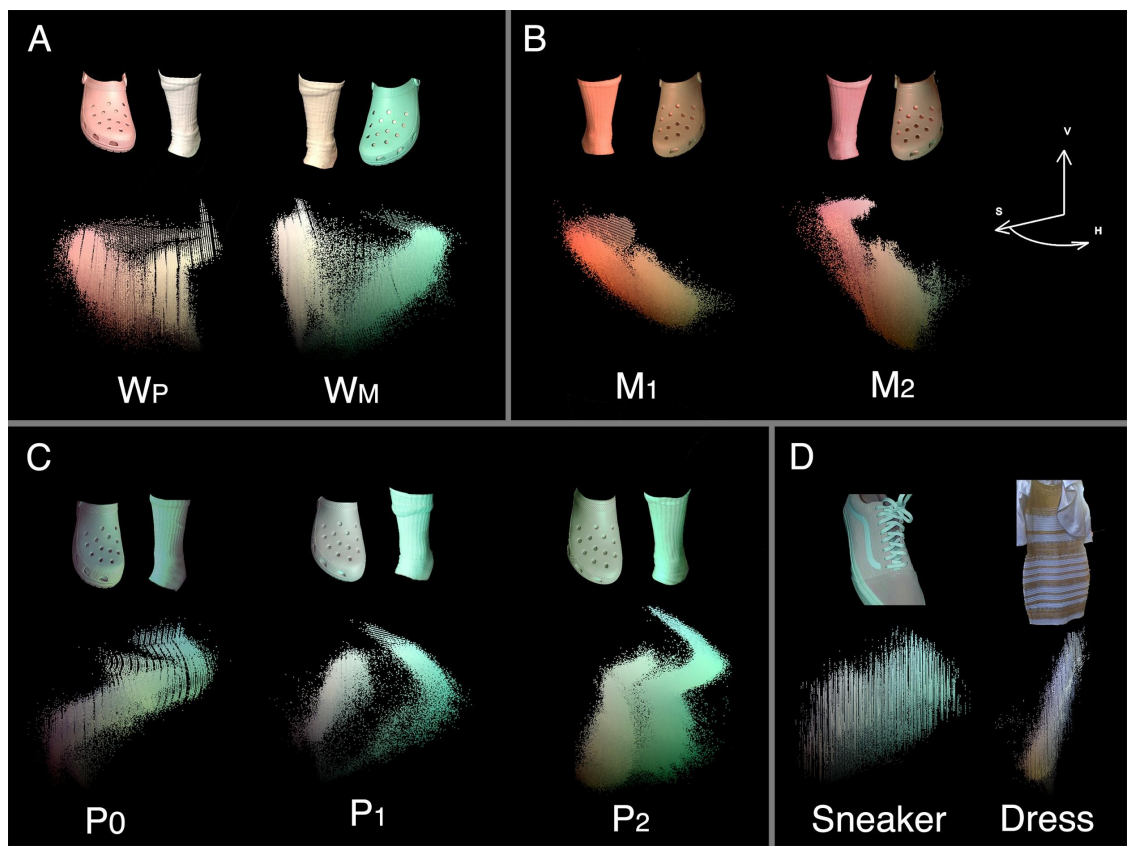

Figure 3: The complete set of stimulus types used in this study. In all cases, we show - for stylistic reasons - a single croc and sock in the top of the croc displays to symbolize the stimulus type. In the study, we used the complete images as depicted in figure 2. The bottom row for each stimulus type depicts all pixels from both crocs and socks (but without black background) as clouds in 2d projections of hsv space, in which individual pixels are represented as colored dots. The axes of hsv space are oriented as depicted in the inset (to the right of $M_2$ in B). These pixel clouds were created using MATLAB's "*colorcloud*" function. In general, the parts in the dot cloud that correspond to the part of the image (croc or sock) are aligned. For example, in P1 the pixels from the sock are to the right and clearly separated from those of the croc. However, this separation is not always as clear, e.g. in M2, where the point clouds blend together, but in general the pixels corresponding to the sock are to the left of those corresponding to the croc. A: "Ballerina pink" (Wp) and "New mint" (Wm) crocs under typical light, as in figure 2A. B: Wm crocs under colored light, as in figure 2B. C: Wp crocs under colored light, as in figure 2C. D: "Sneaker" and "Dress" displays and their corresponding colored pixel clouds. The "Dress" image is courtesy Cecilia Bleasdale.

*Participants*

We logged data from participants online between July 5th and August 5th 2019. These were recruited via social media, specifically Twitter and Facebook. The prompts on Twitter were: "If you have 5 minutes, want to help us out to get to the bottom of why some people see "the crocs" differently and want to see some pretty striking images check this out:" and "What colors do you see, when looking at these #crocs? Last call, if you still want to donate your data. Should only take a couple of minutes.", with a similar prompt on Facebook: "If you have 5 minutes, want to help us out to get to the bottom of why some people see "the crocs" differently and are curious to see some striking ambiguous images, donate your data by taking this brief survey." These invitations yielded data from 5,944 participants. Of these, 108 participants indicated that they were not taking the survey seriously and 80 participants reported a color vision impairment. Thus, we retained data from 5,762 (~97%) participants for further analysis. Of these, 938 (16.3%) identified as male, 3,764 (65.3%) as female and the remainder did not state a gender. The reported median age in our sample was 32 years. All study procedures followed the principles of the Declaration of Helsinki, but were - according to the "final rule" (Revised Common Rule) - exempt from review by the New York University Institutional Review Board (UCAIHS).

*Survey*

We deployed a survey with 18 questions on surveymonkey.com. Nine of these questions pertained to the perception of the stimulus displays. Of these, seven were the crocs and socks displays shown in figure 2. One was "the dress" stimulus and one was a display of "the sneaker", shown in figure 3D. Response options to the croc stimuli were "Pink/salmon", "Green/mint", "Grey/White", "Orange/Beige" and "Other, please specify", ordered randomly for each question. The response options for "the dress" and "the sneaker" stimuli were the standard "white/gold", "black/blue", "blue/gold", "white/black", "switching" and "other" as well as "pink/white", "grey/green", "pink/green", "grey/white", "switching" and "other", respectively. Importantly, as observers could use the other images to calibrate their perception, each image was shown in isolation, on a separate page, with participants unable to go back to a previous page. Participants were prompted with the question: "What color are the crocs in this image?" for croc images, and: "What is the color of the dress in this image?" as well as: "What color is the shoe in this image?" for "the dress" and sneaker images, respectively.

The other questions covered whether observers saw the socks as white, how familiar they are with socks and crocs, what they believed about the source of the illumination and how confident they are in the veracity of their beliefs in general. We also asked demographic questions about age and gender as well as whether observers had a color vision deficit and whether they took the survey seriously.

*Data analysis*

All data were analyzed with MATLAB (Mathworks, Inc., Sherborn, MA). Specifically, we use Monte Carlo methods with 100,000 repetitions to bootstrap confidence intervals (Efron & Tibshirani, 1994). To guard against potential false positives stemming from multiple comparisons, we adopted a conservative confidence interval of 99% for all analyses. Because our data stems from voluntary online survey participation, it is not uncommon that participants did not respond to some questions, e.g. on age or gender. We have no reason to assume that these omissions are systematic, so we adopt a column-wise approach to handling missing data. We report dependent variables in terms of proportions of observers who report a particular percept. As our stimuli are often ambiguous, we report the proportion of our sample with a percept that corresponds to the "Color Appearance Under Typical IlluminatiON" (CAUTION). This avoids the awkwardness of reporting the proportion of participants who report a "veridical" percept, as this concept is ill-defined in the color domain. Veridical pertains to correspondence with reality, but - as has often been noted - while this makes sense for quantities of space and time, the qualitative experience of color is largely generated by the brain and experienced by the mind (Jackson, 1982). Photons have wavelengths, but not colors, and the wavelength of photons is only loosely related to the color experienced by the human or animal observer. Here, "typical illumination" means illumination conditions that observers are likely to encounter in everyday living, such as daylight - when outdoors - or lighting sources designed to mimic daylight when indoors.

**Results**

*Are images of these crocs perceptually ambiguous under normal ("white") lighting conditions?*

The first question that usually arises when discussing profoundly color ambiguous stimuli is whether the disagreement indicates that one of the observers suffers from a color vision deficit. In addition, we have no control over the particular viewing conditions such as viewing distance, screen type, screen calibration, color temperature and the like, which one would be able to control when doing such experiments under laboratory conditions. We cannot even be entirely sure that participants know what a "croc" is or that an image was delivered each and every time, for every participant, given that server uptimes are not perfect. In addition, one could reasonably be skeptical that observers are reliably reporting their percepts in the first place, as we are not paying them nor know their identity nor have any other interactions of any kind.

Thus, it is important to establish that participants can reliably discern and report the color of the croc intended by the manufacturer (pink vs. mint) and that they agree with each other in spite of all of these considerations and possibilities for discrepant reporting of percepts.

We report the results of asking participants about the crocs under typical lighting conditions ($W_P$ and $W_M$) in figure 4.

Remarkably, despite all kinds of theoretical possibilities for misunderstandings and for things to go wrong - whether due to technical issues or participant characteristics of any kind or even interactions between them, about 98% of all participants who provided a response have no problems discerning the manufacturer intended color of the crocs as pink/salmon and green/ mint respectively, from an image served over the internet, and agreeing with each other about the category of the color.

The exact p-value for either outcome obtained with resampling methods is 0, assuming that people used the available color categories at random. At no point did the extremes of the null distribution come anywhere close to the empirical result. Thus, we conclude that the color of these crocs is not ambiguous to our participants under typical lighting conditions.

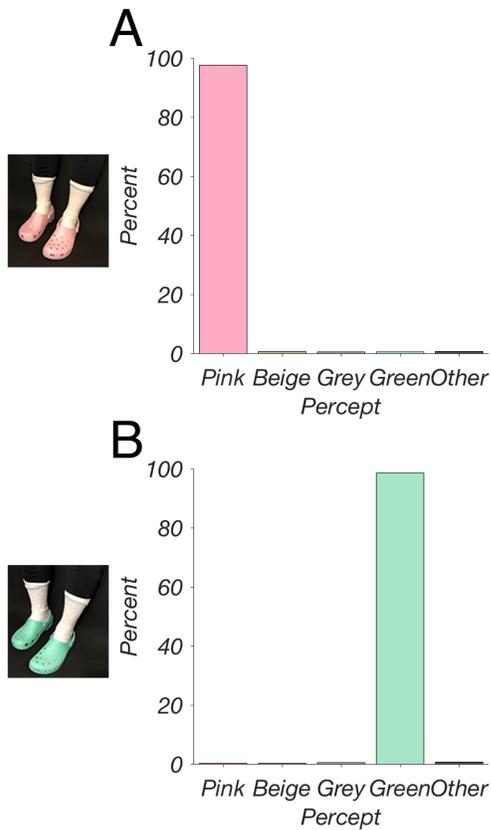

Figure 4: Perception of croc color under typical lighting conditions. A: Percentage of sample reporting categorical percept on the x-axis in response to $W_p$. x-axis: Color labels. For reasons of economy, we abbreviate "pink or salmon" as "Pink", "green or mint" as "Green", "grey or white" as "Grey" and "beige or orange" as "Beige", here and in all other figures. B: As in "A", but for $W_M$. The insets to the left of the y-axes depict the stimulus that evoked the response.

*Are images of these crocs perceptually ambiguous under altered lighting conditions?*

As we saw in figure 4, just putting the crocs on an unpatterned black background does not render them perceptually ambiguous with respect to color, so the question remains whether they do become ambiguous under suitably altered lighting conditions - along the lines we laid out in the Method section.

The results are presented in figure 5. Briefly, we managed to create five categorically color ambiguous stimuli with different perceptual balancing points.

First, we created P0 as a proof of concept. We then created two pairs of stimuli from each colored croc. All 5 images are categorically ambiguous with respect to color - the modal percept for P0 and P1 is grey (43% and 63%, respectively), for P2 green (40%). The M percepts are dominated by beige or orange percepts at 69% (M1) and 61% (M2), respectively. Overall, P2 was the most evenly balanced stimulus - therefore, P2 can be genuinely considered as having four distinct interpretations that are shared by more than 10% of the sample. Also, we note that the available color labels exhausted the space of possibilities well - less than 1% of responses fell into "other" responses.

The chance of observing this pattern of results - using Chi-squared tests with the proportions expected from the croc color perception under normal lighting perceptions is low, with a p-value of below 1e-300 in each of the five cases. Overall, M1, M2 and P1 are conceptually most similar to "the sneaker" (modal percept of grey in about 63% of observers) and "the dress" (modal percept of white/gold in about 60% of observers). Therefore, these stimuli can be considered as bi-stable with respect to the population (not individual observers) and P0 and P2 can be considered tri- and quadra-stable with respect to the population, respectively, in contrast to the unambiguous Wp and Wm.

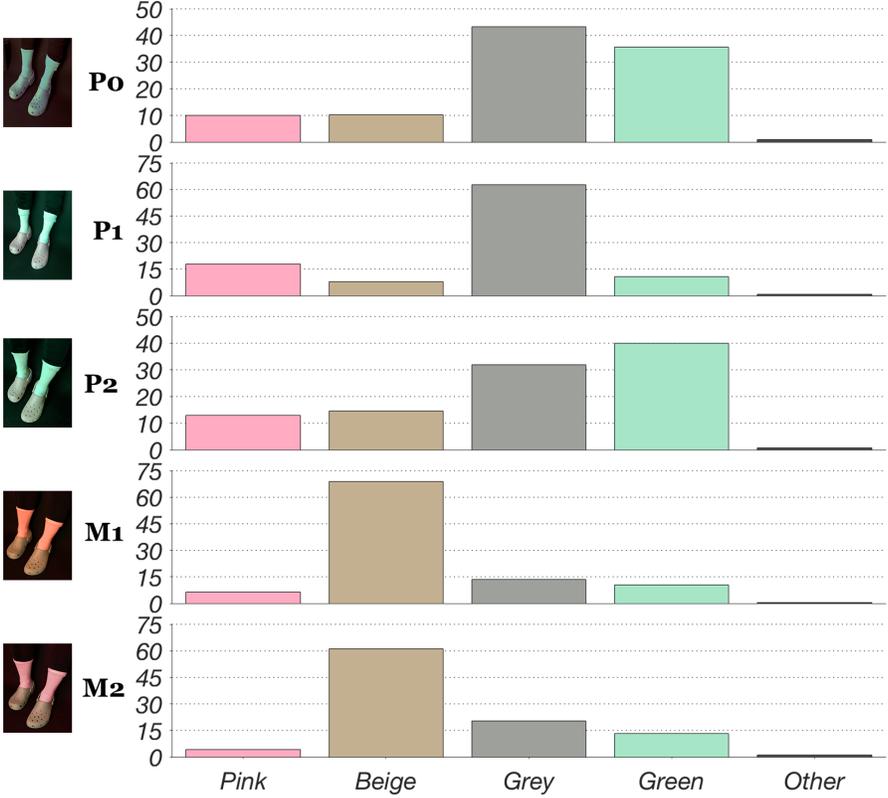

Figure 5: Perception of color ambiguous stimuli created with altered lighting conditions. Rows, from top to bottom: P0, P1, P2, M1, M2, in the order in which we created them. Y-axis: Percentage of sample reporting this percept. X-axis: Categorical color percept. At the beginning of each right, we depict the stimulus that evoked the observer response.

*Are people responding randomly, because participants realize that illumination is not normal?*
Whereas these results are striking at face value, it is conceivable that lighting conditions are so obviously unsuitable that participants cannot reasonably be expected to report their actual percept. Instead, they might be guessing as to what the color of the croc might have been. Such a response could be akin to a situation under low light conditions, when vision is dominated by rod-responses, as rods are more light-sensitive than cones, rendering vision effectively monochromatic (Stabell & Stabell, 2002). However, we don't think this is plausible, for several reasons. First, participants were not shy to point out problems if they arose, and had the option to skip questions. Yet, both the "other" category is almost un-utilized, and most people did respond to the questions. Second, close to 50% of participants did not suspect anything to be amiss with the lighting in the images - guessing that they were either taken under sunlight or indoor lighting, i.e. typical lighting conditions. Finally, and most compellingly, the responses between stimuli are closely related - if someone perceived one image in a certain way, they were far more likely to perceive the others similarly, see figure 6.

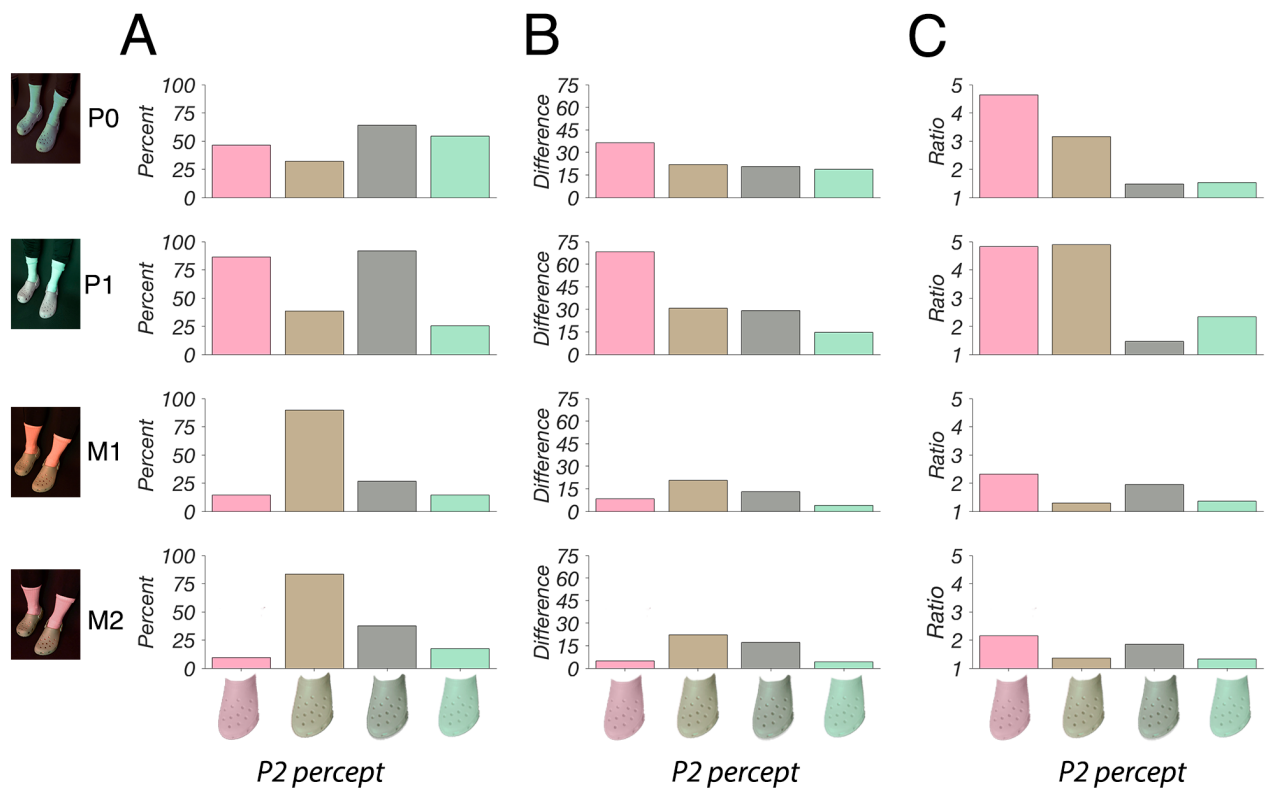

Figure 6: Croc display perceptions as a function of how P2 was perceived. A: In the left panel, we depict the absolute probability of perceiving a croc display in a certain way, as a function of how P2 was perceived (represented by stylized colored crocs on the x-axis). Rows, from top to bottom: P0, P1, M1, M2. B: As in A, but the center panel depicts differences from the baserates depicted in figure 5. C: As in B, but the right panel shows the same information as ratios.

We used P2 perception as a base, because this was the most ambiguous of our stimulus displays. Note that the conditional probability of seeing a particular color is boosted in all 16 cases, sometimes considerably. The average boost was 21 percentage *points*, relative to the percentages shown in figure 5. Sometimes, the result of this boost is dramatic, i.e. the conditional probability of seeing M2 as beige if P2 was seen as beige is 90% (up from 70%). The conditional probability of seeing P1 as pink if P2 was seen as pink is 86% (up from 18%). On average, the ratio is boosted almost 2.5 fold. To summarize, there is considerable consistency within observers and between percepts. This is unlikely to happen accidentally, such as when observers are unsure of which option to pick, particularly as the response choices are randomized and distributed over various pages - importantly - as we noted in the Method section, participants cannot go back and revisit previous stimuli to compare.

Note that the pink percept seems to "gain the most" from having seen the reference stimulus as pink, but this could be simply because this percept had the most to gain to begin with (lowest absolute prevalence in the sample). Also note that this pattern of results - 16 positive differences or 16 ratios larger than 1 - is in itself significant, as we would expect such a result less than 1 in 65,000 cases by chance alone, translating to a p-value of 1.53e-5. As these are conditional probabilities, and the individuals involved different people, these are independent events, so this calculation is valid. Another way to interpret figure 6 is that the P2 percept represents a strong attractor - in whichever way observers perceive P2 - and for whatever reason - increases the probability that other croc displays are seen in the same way. In other words, a propensity to see these stimuli in a certain way cuts across all of these stimuli and might constitute a characteristic of the individual observer.

*What determines the perception of the crocs displays?*
After establishing that participants can agree on the color of the stimuli under typical lighting conditions, but that this consensus breaks down under conditions of SURFPAD lighting, and that this seems to be an individual characteristic, questions about this individual characteristic come into focus. What - specifically - predisposes an individual observer to perceive these croc displays in a certain way - and different from other observers? In the case of "the dress", a key factor turned out to be illumination priors - assumptions about the illumination derived from experience (Witzel et al., 2017; Wallisch, 2017a). We asked observers several questions with this in mind and explore the relative role of fabric and illumination priors here.

Given the previous result - perception of all 5 crocs is tightly inter-related and given that we asked the question about sock color right after asking about P2 - which might have drawn attention to the color of the socks for subsequent stimuli, we focus our discussion of these matters on the percepts involving P2, see figure 7.

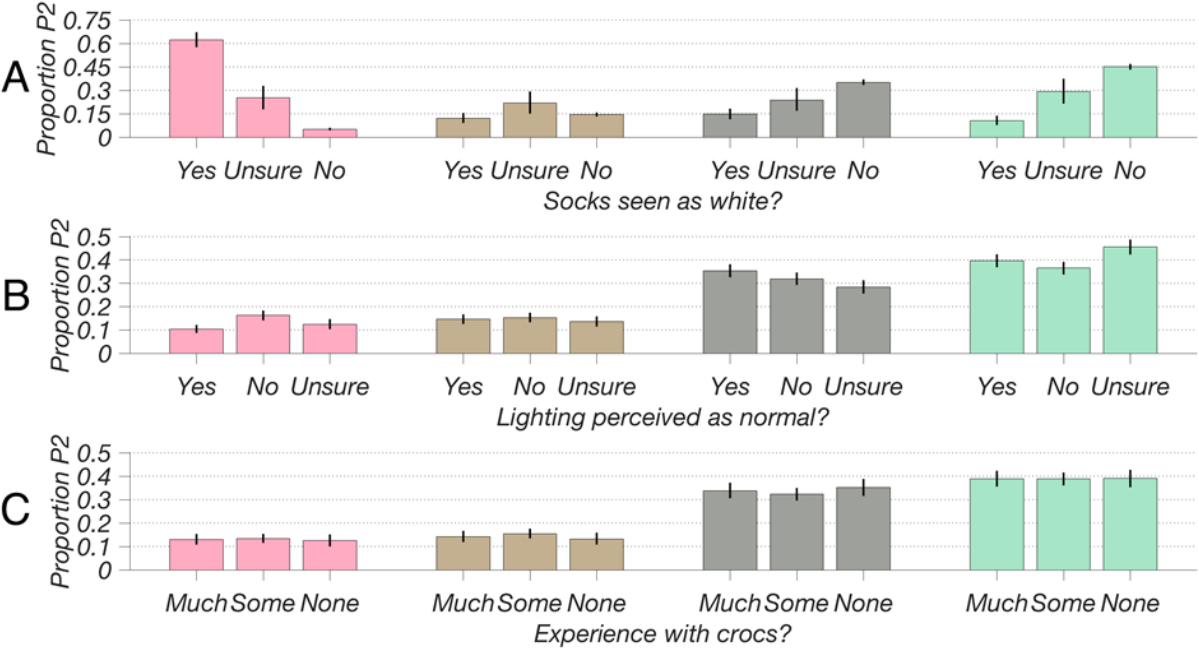

Figure 7: Disagreement about the perception of P2 as a function of responses to relevant survey questions. Each row represents the responses to a self-report question as the independent variable. The dependent variable in all cases is the reported perception of P2. Black vertical bars denote the 99% confidence interval.

The primary driver of P2 perception seems to be how the socks were seen. If the socks were seen as white, the probability that the crocs were perceived as pink is over 12x that of participants who did not perceive the socks as white. Dovetailing nicely with this effect is the observation that the opposite is true for green and grey perception, if the socks were not seen as white, but in the other direction. Observers were more than 2x as likely to perceive the crocs as grey and over 4x as likely to perceive them as green if they did not perceive the socks as white, compared to when they did not. The fact that the confidence intervals are much larger for the "unsure" group is largely due to this being by far the smallest group. A secondary driver of disagreement seems to be lighting conditions. Consistent with our interpretation, if participants did perceive the lighting as abnormal, they were slightly - but significantly - more likely to perceive P2 as pink and less likely to perceive P2 as grey or green, perhaps because they questioned the prima facie appearance of the crocs.

Finally, answers to arbitrary questions do not systematically distinguish participants responding in systematic ways. For instance, we did not expect self-reported croc experience to have any perceptual effects, mostly due to the fact that even being familiar with crocs does not allow to disambiguate which of the 28 classic crocs is on display. In sum, unspecific experience without a clear mechanism does nothing to reduce the uncertainty of the croc color, whereas assumptions about lighting or socks do.

*What determines whether socks were seen as white?*

Along the lines of Wallisch (2017a), we suspect that experience is a primary predictor of assumptions about lighting or sock color, and lifestyle choices a primary predictor of experience. We did not ask a question about lifestyle choices regarding light, as - unlike in "the dress" case, where chronotypes have an obvious impact on sunlight exposure, everything else being equal, there is no clear and reasonably common predictor of exposure to colored LED lights. However, one can ask about direct experience with wearing plain tube socks, so we did ask that question. Figure 8 summarizes the results with regard to the perceptual sequelae of answering this question.

A 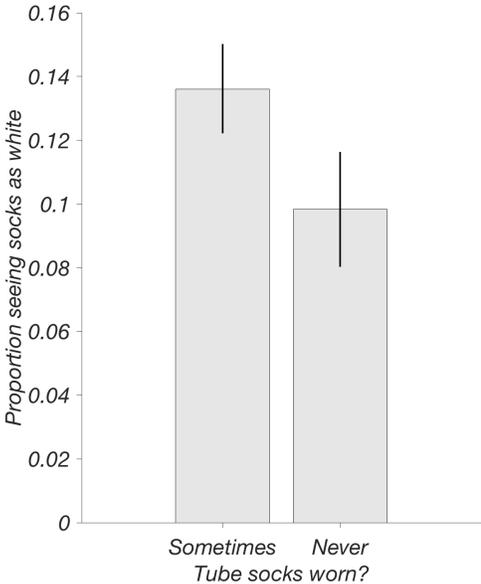 B 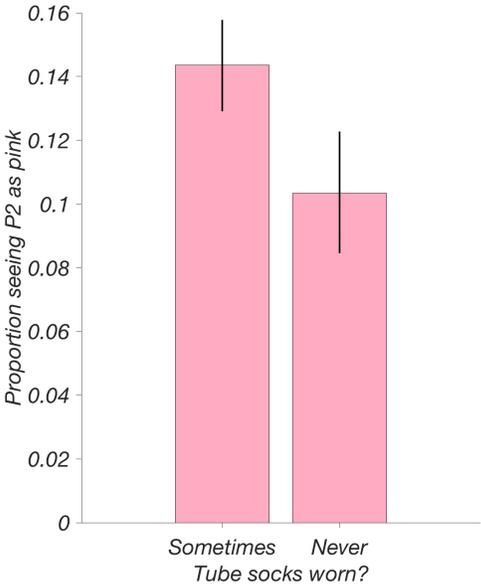

Figure 8: Perceptual consequences of wearing white tube socks. Left panel: Proportion seeing socks as white as a function of tube socks being worn. Right panel: Same, but seeing P2 as pink as the dependent variable. Black vertical bars represent the 99% confidence interval.

The effects in figure 8 are obviously less strong than what would be needed to comfortably claim that we understand this aspect of the phenomenon well, but then again it is both remarkable and interesting to see these systematic and significant effects in the first place, given how much random variability is likely involved in these self-reported responses. In addition, there are certainly other ways to be exposed to tube socks than wearing them, i.e. one might see them in sports, on TV, in ads, or in stores, but none of these are easy to probe in self-reports. Put differently, we would not expect strong effects on the basis of this simple self-report question, as wearing the socks is clearly not a very close proxy to actual overall exposure. Interestingly, the perceptual effects on P2 trade off clearly, i.e. all of the increased pink perception comes from a decrease in perceiving P2 as green, with the proportions of grey and beige entirely unaffected (not pictured in figure 8), just as one would expect if the socks are reflecting the light, allowing observers to discount it.

*How broad is the effect of these priors on perception more generally?*
Now that we established the power of these priors on the perception of P2 with respect to color, one can wonder how far these priors extend. Does P2 perception predict perception of "the sneaker" and "the dress"? An answer in the affirmative could hint at common underlying mechanisms and observer characteristica. Does a strong sock prior also predict perception of "the sneaker" or "the dress"? For answers to these questions, see figure 9.

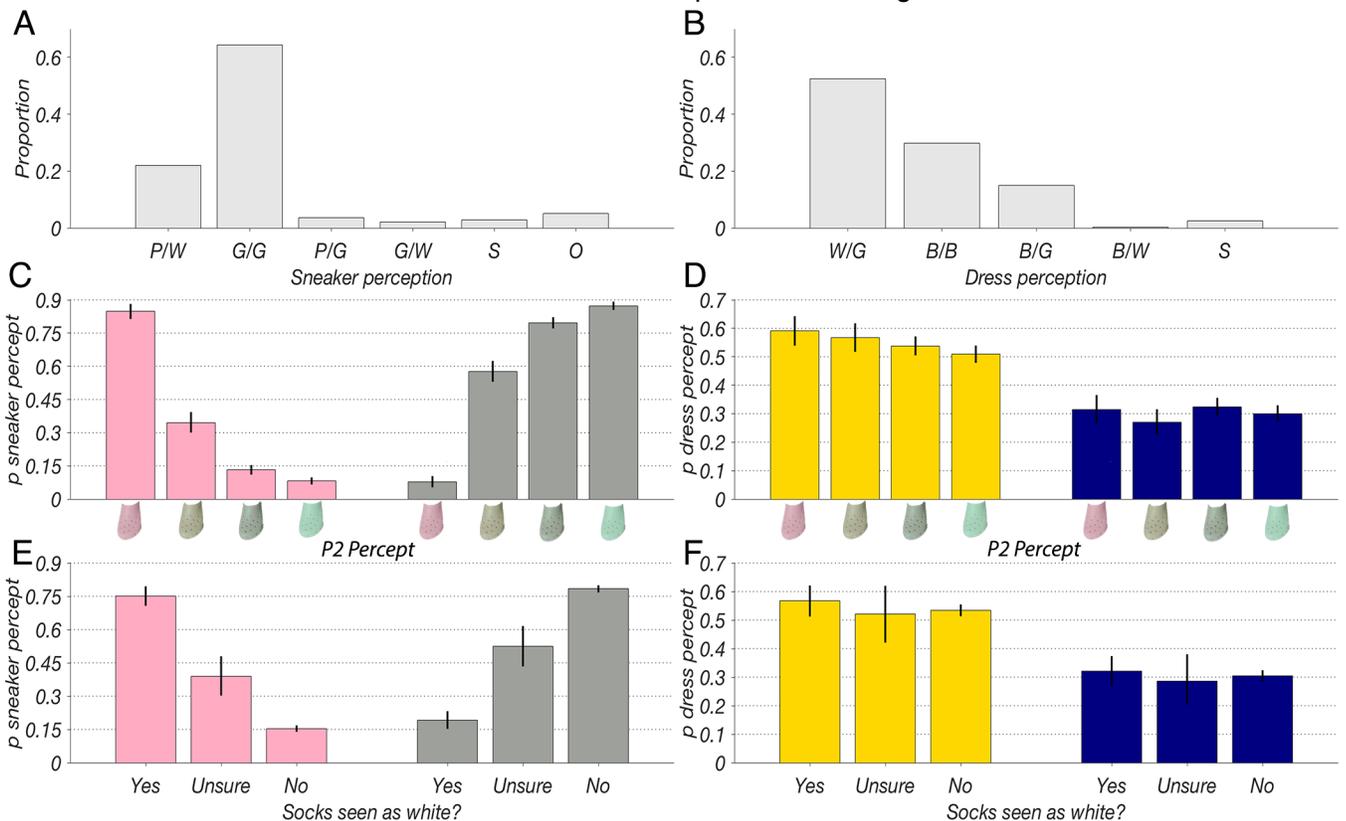

Figure 9: Priors and non-croc perception. A: Descriptive statistics for sneaker perception. Y-axis: Proportion of the sample reporting seeing Pink/White (P/W), Grey/Green (G/G), Pink/Green (P/G), Grey/White (G/W), Switching (S) or Other. B: Descriptive statistics for dress perception. Y-Axis: Proportion of the sample reporting seeing White/Gold (W/G), Blue/Black (B/B), Blue/Gold (B/G) and Black/White (B/W) or Switching (S). C: Probability of seeing the sneaker as pink (pink bars) or grey (grey bars) as a function of seeing P2 as pink, green, grey or beige (as indicated by the croc pictograms on the x-axis). D: Probability of seeing the dress as White/Gold (Yellow bars) or Black/Blue (Blue bars) as a function of seeing P2 as pink, green, grey or beige (as indicated by the croc pictograms on the x-axis). E: Probability of seeing the sneaker as pink (pink bars) or grey (grey bars) as a function of sock perception. F: Probability of seeing the dress as White/Gold (Yellow bars) or Black/Blue (Blue bars) as a function of sock perception. Black vertical bars represent the 99% confidence interval.

The first thing to note is that "the sneaker" is most perceptually similar to P1 - grey perception strongly dominates, with the remainder mostly taken up by pink, as could be anticipated from the point clouds in figure 3. Second, we broadly replicate the proportions described in the perception of "the dress" as reported in Lafer Sousa et al. (2015), or Wallisch (2017a), even four years later. Third, P2 perception strongly determines sneaker perception. Almost 90% of those who saw P2 as pink saw the sneaker as pink, whereas just over 10% of those who saw P2 as green did. Almost 90% of those who saw P2 as green saw the sneaker as grey, whereas just over 10% of those who saw P2 as pink did. The numbers for seeing P2 as grey is almost the same as seeing P2 as green, whereas seeing P2 as beige falls somewhere in between grey and pink. In contrast, P2 perception had little to no influence on the perception of "the dress", suggesting that crocs and sneaker perception share a common mechanism - perhaps fabric priors, specifically white sock priors - whereas the perception of "the dress" is primarily driven by a different mechanism - perhaps lighting priors, specifically sunlight priors. Fourth, the effects of assuming the socks are white reach across percepts, predicting not only P2 - and indeed the perception of the other croc displays - but also the sneaker. This could be due to a number of reasons. For instance, this could be a manifestation of a tendency to apply fabric priors across the board. Or this pattern of results could simply reflect the fact that people who think socks are white are the same people who think that shoelaces - and maybe other objects like that - are white. It could also reflect the workings of another, not yet well understood mechanism or personal characteristic, of which sock perception is simply a marker of. Notably, these socks are often worn in athletic contexts - it would not be too far of a stretch to suggest that they might be worn with - sneakers - that will tend to happen to have white shoelaces. Finally, consistent with what we already observed here, how the sock is perceived seems to have very little impact on how the dress is perceived, again suggesting that the perception of "the dress" is not governed by fabric priors, but primarily by lighting priors. In addition, the lack of an impact of sock perception on dress perception suggests that there isn't simply some overall bias in color perception that affects all of perception affecting the color perception of socks, crocs, sneakers and dresses alike. The complete lack of an impact on dress perception rules out this possibility - priors and how they manifest in perception seem to be fairly specific.

*What is the cognitive penetrability of these effects?*

Most perceptual effects are notoriously immune to awareness and cognitive interventions (Firestone & Scholl, 2016). Telling people that the physical "dress" is blue and black does not change one's percept, should it be white and gold. However, there is reason to believe that the effect driven croc perception might be different. For instance, if one were to take one's percept at face value, a pink croc under typical lighting conditions, as in Wp would likely be perceived as grey or green under altered lighting conditions, as these are the colors of the pixels that make up the croc (see figure 3). However, if one were to recognize that these lighting conditions are irregular, one might scrutinize them and give more weight to one's priors, importantly one's white sock prior, if such a prior exists, allowing an observer to color calibrate the image and perhaps recover the color under typical light. To account for this possibility, we asked one question along these lines, namely whether observers consider it possible that their beliefs could be wrong. This question is somewhat difficult to interpret, as a lot of factors might go into answering it, including self-awareness, meta-cognition, modesty, as well as unequal efforts among people to make sure that their beliefs are accurate. In hindsight, we should have asked something more closely related to perception, i.e. beliefs in naive realism, but we do think it is fair to assume that this question can be taken as a proxy for confidence in the veracity of one's own cognition, given everything else is equal. Of course, everything else won't be equal on the level of the individual, which is why a large sample size is critical. See figure 10 for a summary of these effects.

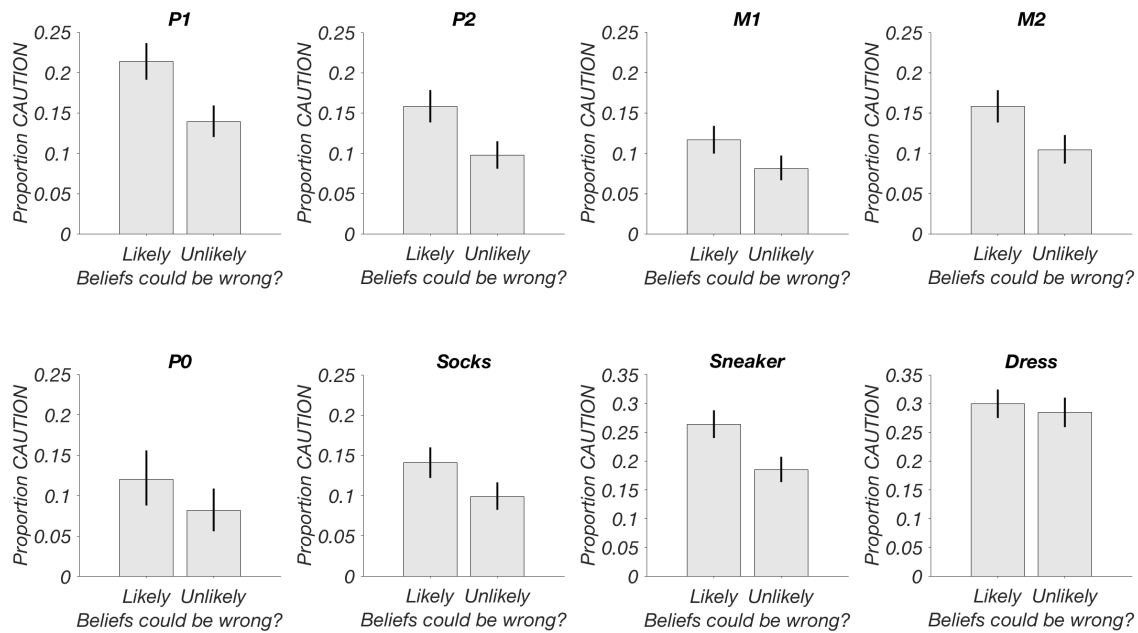

Figure 10: Cognitive penetrability of percepts. In each panel, we report the proportion of participants who report a color perception that corresponds to the color appearance under typical illumination conditions (CAUTION, see Method) as a function of whether participants allowed for the possibility that some of their beliefs might be wrong, for a given stimulus display. Black vertical bars represent the 99% confidence interval.

The effects illustrated in this figure are consistent and perhaps even ironic. The more confident observers are in their beliefs being correct, the less likely they are to recover CAUTION, which is not implied by the pixel clouds in figure 3. In other words, these observers trust the appearance of the shoes, without second-guessing this appearance, perhaps due to confidence in the general veracity of one's cognition. This extends to CAUTION perceptions of socks and sneakers, which are significantly modulated by the answer to this question. As we noted above - the perception of "the dress" is not influenced by meta-cognition. We have confidence that the confidence intervals in $P_0$ would shrink considerably with more participants - we had by far the least respondents to this question. Whether this is ironic or not - echoing other meta-cognitive effects (Dunning, 2011) depends on one's conception of what color the crocs "really" are - CAUTION or the color of the individual pixels as seen in the point clouds of figure 3, or if that question even makes sense, as the concept of veridicality is not well defined for color in the first place.

*What is the effect of demographics?*
Finally, we wondered about the effects of demographics on the perception of these stimuli. Prior research on "the dress" (e.g. Lafer-Sousa et al., 2015; Wallisch, 2017a) suggest that gender effects are miniscule, but age effects could be considerable, although a relative dearth of old observers is a concern. See figure 11 for an overview of these demographic effects.

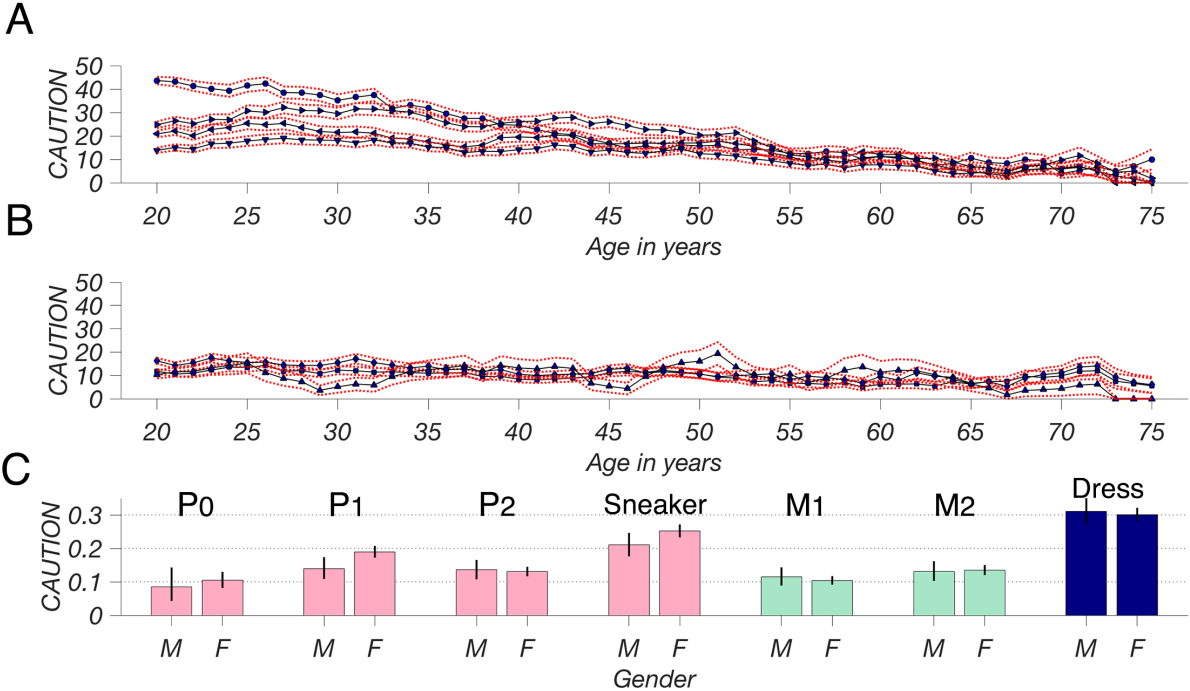

Figure 11: Demographic effects. A: Stimuli where CAUTION decreases with age. Circles: Dress. Right-pointed triangles: Sneaker. Left-pointed-triangles: P1. Down-pointed triangles: P2. Red dotted lines represent SEM. B: Stimuli where there is no clear age-effect. Squares: M1. Diamonds: M2. Up-pointed triangles: P0. Red dotted lines represent SEM. C: Effects of gender

on CAUTION. Each pair of bars corresponds to the stimulus displays labeled on top of it. Black vertical bars represent the 99% confidence interval.

To summarize, most pink croc stimuli, as well as the dress and sneaker exhibit strong age-effects, with a gradual decrease in CAUTION as a function of age. This contrasts with a sharp decline of the White/Gold modal percept with retirement age for the dress (Wallisch, 2017a), which we replicated here and a sharp rise in Blue/Gold percepts (not pictured). As noted in Wallisch (2017a), it is hard to meaningfully interpret these effects, as age is a carrier variable of so many other effects, including physiological effects and cohort effects. We also observe a lack of an age related decline in CAUTION for the mint croc stimuli, which could possibly be due to a floor effect - there is little CAUTION for these stimuli to begin with. We attribute the lack of an age effect for P0 to a lack of participants, as in figure 10 - the error bars are simply too wide to interpret this trajectory meaningfully. Finally, as already discussed, gender effects are miniscule, if they exist at all. There seems to be marginal gender effect on the perception of P1 and its related stimulus, the sneaker, but if they exist, these effects are small and hard to interpret. This addresses one of the concerns related to this study, namely that the majority of our sample identified as female. This might be owed to the fact that we framed the recruitment for our study in terms of donating data, and women have been shown to be more proactive in charitable giving in general (Piper & Schnepf, 2008; Mesch et al., 2011). As figure 11 illustrates, whereas this might have biased the gender-composition of the sample, it is unlikely that this imbalance affected the perceptual effects reported in this study.

**Discussion**

We managed to create visual displays such that observers categorically disagree about the colors of the objects depicted, even though the same objects evoke near complete consensus under typical lighting conditions. We did so by systematically avoiding, minimizing and reducing cues that observers typically use to determine the color appearance of an object, such as patterned backgrounds (Jenness & Shevell, 1995), color memory (Witzel et al., 2011) or familiar lighting (Maloney, 1999), while at the same time introducing a part of the object that mirrors the properties of the lighting, but for which some - but not all - participants have a prior that they use to anchor their color percept, a process which we term SURFPAD. We were also able to show that the perception of the different color displays was systematically linked within a given observer, extending even to other color-ambiguous stimuli, such as the "sneaker". We take this to mean that a disposition to see these stimuli in a certain way amounts to a personality characteristic. This personality characteristic is - in this case - best characterized as a "fabric prior" for white socks, which strongly determines the perception of the displays we created. This fabric prior - in turn - seems to have been determined by experience with this type of sock, akin to the illumination priors that determine the perception of the "dress" due to differential exposure to sunlight (Wallisch, 2017a).

We believe that these effects are consistent with the operation of known cognitive mechanism involved in color vision, most importantly color constancy (McCann et al., 1976). In this special case, color constancy seems to be derived from a kind of color memory (Jin & Shevell, 1996; Hansen et al., 2006), in a subset of our observers. Remarkably, this mechanism allows these observers to recover the color that corresponds to CAUTION, even though there are virtually no individual pixels in the object that suggest CAUTION, a striking empirical instance of the whole being different from the sum of its parts, as suggested theoretically by Koffka (1922). Put differently, instead of alerting the organism to the profound uncertainty inherent in these displays, the brain seems to err on the side of arriving at actionable conclusions by operating inadvertent mechanisms akin to autocorrect - educated guesses to recover the stimulus situation by making assumptions about the color of lighting (in the case of the dress) or fabric (here). It is increasingly well recognized that perceptual mechanisms in the brain utilize information beyond the present stimulus in the form of priors, assumptions, expectations and predictions, in order to cope with and act in an environment that is inherently more uncertain than one would prefer (Press & Yon, 2019). This line of reasoning also suggests that while we understand the color appearance of objects under typical lighting conditions relatively well - to the point of being able to reasonably describe color appearance in terms of a linear system (Shevell, 2003), the same might be true under carefully altered lighting conditions, where differential priors play a much larger role in determining the subjective color experience.

There are several limitations inherent in this study that potentially threaten the generality of these conclusions. Most of these pertain to our specific implementation. First, in order to rapidly obtain data from many observers, we logged the data online, instead of in a lab. This will necessarily introduce considerable uncertainty as to the stimulus situation that the participants actually encountered, such as screen size or color calibration, which would not be a concern under controlled laboratory conditions. Similarly, we would have wanted to take more complete control of the lighting conditions when creating these stimuli, involving both a high-end LED lighting system and a photometer. Then again, that we readily found these effects despite the noise introduced by presenting the stimuli online speaks to both their strength and robustness, akin to all the other color-ambiguous stimuli (the "dress", "adidas jacket", "flip-flops", etc.) that surfaced and thrived "in the wild", without relying on carefully controlled laboratory conditions. In addition, our observers were able to agree on the color of these objects under typical lighting conditions, so it is unlikely that this ambiguity is introduced as an artifact of different screen settings. Similarly, whereas we strongly suggest future studies in this domain to more carefully control the lighting conditions when creating the ambiguous displays, our efforts still have merit as a proof of concept. What seems to matter critically is the relative position of the color clouds from the two objects - the colored target object (CTO), the croc in this case and the cueing or anchoring illumination-mirroring object (IMO), in this case the sock - in hsv space. Displays seem to evoke particularly strong disagreement between observers if there is a separation between the two clouds and if they form parallel columns in this space, whether this configuration is achieved by a careful control of lighting conditions or digital filtering. The creation of such stimuli in hsv space will also allow us to address the issue of beige perception. Are participants perceiving these stimuli as beige because they tried to color-correct their percept, but failed (ending up at beige instead of pink), or are the actual pixels - as suggested by the point clouds in figure 3, particularly for M1 and M2 far enough in the beige range to make a beige croc color plausible a priori? Cleaner stimuli - that are clearly separated and mostly avoid the beige range should cut down on beige perception, increasing the perception of pink (for the Ps) and grey (for M).

Another concern is that we illustrated the SURFPAD principle only with one type of object - crocs and socks, while claiming that it is a general design principle. This is a clear case of a concern that can readily be addressed in future studies, using any number of colored stimuli, such as colored candles as the CTO and the wick as the IMO or colored tissue boxes as the CTO and white tissues as the IMO. With some imagination, this principle can likely be extended to other visual domains, such as distance or motion, and perhaps even to auditory stimuli. Based on our experiences of applying SURFPAD to crocs and socks, we predict that a suitable application of these principles would yield stimulus displays with similar levels of disagreement between observers.

However, the biggest concern regarding this kind of work is that we overlooked some low level explanation for the phenomena we observed, like so many others in perception (Firestone & Scholl, 2016). It would not be the first time that a high profile and high level explanation is better explained by other factors, which is usually the case when the experimenter inadvertently fails to notice or rule out covariates that correlate and drive the dependent variable even better. A prominent example of this would be Mischel who thought that willpower determines the ability to resist the temptation of marshmallows (Mischel et al., 1989), a phenomenon better explained by trust (Watts et al., 2018). This can happen even in experimental work, when the independent variables created by the experimenter mask a confound. For instance, it has been proposed that observers perceiving a rotating line to lead a simultaneously flashed line suggests the operation of motion extrapolation mechanisms to guide dynamic action (Nijhawan, 1994). However, this phenomenon can be more readily explained in terms of a low-level feature - the flashed line has lower effective contrast than the rotating one, and lower contrast is associated with longer neural delays (Raiguel et al., 1999). This suggests that the rotating line is seen as leading because its signal arrives in the brain before that of the flashed line, so it should be possible to negate this advantage - and the lead - by increasing the relative contrast of the flashed line, which is what can be shown empirically (Purushothaman et al., 1998).

While we concede this theoretical possibility, we do not think it likely in this case. First, the SURFPAD principle leverages a combination of known mechanisms and principles of visual processing, without relying on exotic postulates or remote theoretical possibilities. Second, given how rare such color ambiguous displays are - with the advent of smartphones, there are now well over a trillion images taken per year (Richter, 2017), and only a handful of these are known to be color-ambiguous, whereas all the displays we created according to this principle are color ambiguous, creating such displays inadvertently is about as plausible as suggesting to go to the moon without a solid understanding of gravity or to build a nuclear reactor without a reasonable understanding of the basic physics governing nuclear reactions involved.

Finally, suggesting that there are more powerful low level features at work that we somehow overlooked strains credulity, given the sheer size of the effects at play - the white sock prior alone yields a 12-fold range in the pink percept of P2, which in turn determines close to 90% of the sneaker percept. One would be hard-pressed to envision more powerful predictors of these percepts.

Of course, it is true that - in principle - a final determination on this issue won't be possible without future work. If we claim that socks are critical to bring about disagreement of the croc color, it stands to reason that we should create and ask observers about crocs altogether without socks. As this might look somewhat strange, a fair comparison would be to use the same exact crocs, but making sure that the socks do not reflect the light. Using perfectly white socks should not allow observers to use the socks to color calibrate the rest of the display. Similarly, it might be of interest to ask observers directly about the color of shoelaces to further elucidate the link between sneaker perception and croc perception, and the role of fabric priors and their specificity as they apply to perception.

Our results do suggest that disagreements due to differential illumination priors (as in the dress) and disagreements due to differential fabric priors (as in the crocs) are similar, but separate phenomena. We believe that this is due to the fact that the "dress" phenomenon is situated on the daylight axis between yellow and blue, whereas our crocs displays leverage the opposite axis between pink and green, which is likely why there is so little cross-display information. In other words, we can't predict how someone saw the dress based on how they saw the crocs or vice versa, because these displays operate on different axes through color space. We propose to create a display involving both fabric priors and daylight priors involving grey, yellow and blue crocs as well as yellow and blue lights, using SURFPAD principles.

A bigger point of this study - beyond crocs, socks and color - is that this principle pertains to the nature of disagreement and polarization. Given some of the feedback we received, there is no question that some people are annoyed by these kinds of ambiguous displays, as being confronted with these displays threatens their self-perception as being accurate observers of and effective operators in the external world. Of course, this kind of self-perception is validated all the time - under typical conditions, people's interpretation of the situation does correspond to reality, and the interpretation of others. What makes the operation of these cognitive "autocomplete" or "autocorrect" mechanisms so pernicious is that they do not flag to the observer when they had to be leveraged to bring about the perceptual interpretation. They are simply presented to the observer as "the percept", regardless of how it came about, perhaps in an attempt to facilitate and encourage action.

Note that we would expect these effects to be least worrisome in perception, for several reasons. First, the brain devotes considerable resources to the processing of perceptual information (Wallisch & Movshon, 2008). Second, there are - under natural conditions - multiple redundancies both within and across modalities (Landy & Kojima, 2001; Hillis et al., 2002; Körding et al., 2007). Third, there was likely immense evolutionary pressures to get the right answer. If the brain is indeed epistemically immodest - biased towards deriving actionable conclusions despite profound uncertainty - as suggested by our results, it seems important that these conclusions are not catastrophic to the integrity of the organism. So whereas these mechanisms are particularly tractable in the perceptual domain, we suspect that they play a larger role in cognition beyond perception, particularly as we now live in environments and systems that are complex enough that there are almost always considerable degrees of freedom for observers to come up with their own interpretation (Wallisch & Whritner, 2017). Put differently, we are now - cognitively - living in a SURFPAD world. We are increasingly encountering conditions of tremendous and polarized differences in the interpretations of current events. This is somewhat surprising, as it is happening in spite of all the information that is now available about the state of the world, which could drive consensus. But it is important to note that given the complexity of the world, we are almost always in a regimen of high SU (due to the complexity of reality and the social systems we inhabit itself). So if there is fundamental disagreement (D), it is likely not due to disagreements about the evidence - shrouded in SU - but rather the cognitively invisible RFPA. Where do these ramified or forked priors and assumptions come from?

It is no secret that media and social media have effectively created increasingly divergent differential priors, leading to starkly differing interpretations of events, regardless of what actually happens. Alarmingly, both media and social media seem to have created cognitive versions of the socks, anchors that appear politically correct to a particular echo chamber (Brady et al., 2019). To explore reductions in the difference in interpretations, future work on SURFPAD could explore the malleability of these differential priors - does priming with suitable stimuli just before the study update the prior? Is it possible to override the prior by introducing other display elements that could restore consensus, e.g. a colored background that also reflects the illumination?

Thus, in terms of fostering conflict resolution, we advise neither to highlight the disagreement itself - which typically garners all the attention, nor the SU, which is inevitable in a complex world, but rather to focus squarely on the roots of disagreement - the RFPA. Differential priors, created in a strong and polarized fashion and under conditions of profound uncertainty will make dramatically differing interpretations of the same events and situations inevitable.

It is appropriate to address this issue with the urgency that it necessitates if one is to avoid the adverse outcomes associated with mounting and polarized disagreement.